\documentclass[preprint,preprintnumbers,showkeys,amsmath,amssymb,superscriptaddress]{revtex4}
\usepackage{txfonts}% Physical Review B
\usepackage{graphicx}% Include figure files
\usepackage{dcolumn}% Align table columns on decimal point
\usepackage{bm}% bold math
\usepackage{array}
\usepackage[colorlinks=true,plainpages=false,linkcolor=blue,urlcolor=blue,citecolor=blue,pdfpagemode=UseNone,pdfstartview=FitBH]{hyperref}

\begin{document}

\title{Microscopic evidence of charge- and spin-density waves in La$_3$Ni$_2$O$_{7-\delta}$ revealed by $^{139}$La-NQR}

\author{J. Luo}
\thanks{These authors contributed equally to this work.}
\affiliation{Institute of Physics, Chinese Academy of Sciences,\\
 and Beijing National Laboratory for Condensed Matter Physics,Beijing 100190, China}
\affiliation{School of Physical Sciences, University of Chinese Academy of Sciences, Beijing 100190, China}

\author{J. Feng}
\thanks{These authors contributed equally to this work.}
\affiliation{Institute of Physics, Chinese Academy of Sciences,\\
 and Beijing National Laboratory for Condensed Matter Physics,Beijing 100190, China}
\affiliation{School of Physical Sciences, University of Chinese Academy of Sciences, Beijing 100190, China}

\author{G. Wang}
\affiliation{Institute of Physics, Chinese Academy of Sciences,\\
 and Beijing National Laboratory for Condensed Matter Physics,Beijing 100190, China}
\affiliation{School of Physical Sciences, University of Chinese Academy of Sciences, Beijing 100190, China}

\author{N. N. Wang}
\affiliation{Institute of Physics, Chinese Academy of Sciences,\\
 and Beijing National Laboratory for Condensed Matter Physics,Beijing 100190, China}
\affiliation{School of Physical Sciences, University of Chinese Academy of Sciences, Beijing 100190, China}

\author{J. Dou}
\affiliation{Institute of Physics, Chinese Academy of Sciences,\\
 and Beijing National Laboratory for Condensed Matter Physics,Beijing 100190, China}
\affiliation{School of Physical Sciences, University of Chinese Academy of Sciences, Beijing 100190, China}

\author{A. F. Fang}
\affiliation{School of Physics and Astronomy, Beijing Normal University, Beijing 100875, China}
\affiliation{Key Laboratory of Multiscale Spin Physics, Ministry of Education, Beijing Normal University, Beijing 100875, China}

\author{J. Yang}
\affiliation{Institute of Physics, Chinese Academy of Sciences,\\
 and Beijing National Laboratory for Condensed Matter Physics,Beijing 100190, China}
\affiliation{School of Physical Sciences, University of Chinese Academy of Sciences, Beijing 100190, China}

\author{J. G. Cheng}
\affiliation{Institute of Physics, Chinese Academy of Sciences,\\
 and Beijing National Laboratory for Condensed Matter Physics,Beijing 100190, China}
\affiliation{School of Physical Sciences, University of Chinese Academy of Sciences, Beijing 100190, China}

\author{Guo-qing Zheng}
\affiliation{Department of Physics, Okayama University, Okayama 700-8530, Japan}

\author{R. Zhou}
\email{rzhou@iphy.ac.cn}
\affiliation{Institute of Physics, Chinese Academy of Sciences,\\
 and Beijing National Laboratory for Condensed Matter Physics,Beijing 100190, China}
\affiliation{School of Physical Sciences, University of Chinese Academy of Sciences, Beijing 100190, China}

\date{\today}% It is always \today, today,

\begin{abstract}
The recent discovery of superconductivity in La$_3$Ni$_2$O$_{7-\delta}$ with a transition temperature $T_c$ close to 80 K at high pressures has attracted significant attention, due particularly to a possible density wave (DW) transition occurring near the superconducting dome. Identifying the type of DW order is crucial for understanding the origin of superconductivity in this system. However, owing to the presence of La$_4$Ni$_3$O$_{10}$ and other intergrowth phases in La$_3$Ni$_2$O$_{7-\delta}$ samples, extracting the intrinsic information from the La$_3$Ni$_2$O$_7$ phase is challenging. In this study, we employed $^{139}$La nuclear quadrupole resonance (NQR) measurements to eliminate the influence of other structural phases in the sample and obtain microscopic insights into the DW transition in La$_3$Ni$_2$O$_{7-\delta}$. Below the DW transition temperature $T_{\rm DW} \sim$ 153K, we observe a distinct splitting in the $\pm$ 5/2 $\leftrightarrow$ $\pm$ 7/2 transition of the NQR resonance peak at the La(2) site, while only a line broadening is seen in the $\pm$ 3/2 $\leftrightarrow$ $\pm$ 5/2 transition peak. Through further analysis of the spectra, we show that the line splitting is due to a unidirectional charge modulation. A magnetic line broadening is also observed below $T_{\rm DW}$, accompanied by a large enhancement of the spin-lattice relaxation rate, indicating the formation of magnetically ordered moments in the DW state. Our results suggest a simultaneous formation of charge- and spin-density wave order in La$_3$Ni$_2$O$_{7-\delta}$ , thereby offering critical insights into the electronic correlations in Ni-based superconductors. 
\end{abstract}

%\pacs{74.25.nj, 74.40.-n, 74.25.Dw}

\maketitle

The mechanism of the cuprate high temperature superconductors is still under debate since their discovery\cite{Cuprate,cuprateReview}. 
%New superconductor with high critical temperature \emph{T}$_{\rm c}$   The attention on nickelates arises since 1950s before the discovery of cuprate\cite{Cuprate}. 
%However, the going was slow in the past thirty years. In 2019, superconductivity with \emph{T}$_{\rm c}$ = 9 - 15 K was found in infinite-layer  Nd$_{0.8}$Sr$_{0.2}$NiO$_{2}$ thin films\cite{nickelatethinfilm}, and \emph{T}$_{\rm c}$ reaches to 30K under high pressure in related compound  Pr$_{0.82}$Sr$_{0.18}$NiO$_{2}$\cite{Nifilmhighpressure}. 
Recently, La$_3$Ni$_2$O$_{7-\delta}$ shows superconductivity with the superconducting transition temperature $T_{c}$ close to 80 K at $P >$ 14 GPa\cite{327Nature,327CPL1}, generating a tremendous response from the research community and providing a new platform to study the mechanism of high temperature superconductivity\cite{327CPLreview}. La$_3$Ni$_2$O$_{7}$ (327) belongs to Ruddlesden-Popper (R-P) phases La$_{n+1}$Ni$_{n}$O$_{3n+1}$ with $n$ = 2, including two NiO$_6$ octahedral layers separated by La-O layer in the unit cell. %The space group of La$_3$Ni$_2$O$_{7-\delta}$ is $Amam$ at ambient pressure, crystallized to orthorhombic structure.
%Surprisingly, when the pressure increases to 14 GPa in single crystal La$_3$Ni$_2$O$_{7-\delta}$, the resistivity exhibits a distinct drop at $T = $ 78.2 K, perhaps a signature of superconductivity\cite{327Nature,327CAC}. 
Within the pressure range where superconductivity emerges, the structure concurrently changes from orthorhombic to tetragonal as revealed by X-ray diffraction (XRD)\cite{327Nature,327I4mmm}. %Subsequently, in situ synchrotron XRD experiments uncover a tetragonal phase $I4/mmm$ around 19GPa at 40K\cite{327I4mmm}, possibly the structure where superconductivity occurs. 
Later, both the single crystal and polycrystalline samples achieve zero resistance at low temperatures\cite{singlecrystalSC,polycrystalsol-gel}. However, due to the presence of La$_2$NiO$_{4}$ (214) and La$_4$Ni$_3$O$_{10}$ (4310) phases, as well as the intergrowth of different R–P phases observed by scanning transmission electron microscope (STEM) and nuclear quadrupole resonance (NQR), the superconducting volume fraction is found to be less than 1$\%$ in La$_3$Ni$_2$O$_{7-\delta}$\cite{filamentarySC,327LaPrNiOnature}. With Pr doping, both 4310 and intergrowth phases can be strongly suppressed, and the superconducting volume fraction can even exceed 90$\%$\cite{327LaPrNiOnature}.

Not only the high transition temperature $T_c$, but another  feature  that a possible density wave (DW) state is adjacent to the superconducting phase is intriguing. This was initially observed through the abrupt change in the resistivity, Hall coefficient and Seebeck coefficient at $T \sim$ 120 K in La$_3$Ni$_2$O$_{7-\delta}$ under ambient pressure\cite{Sreedhar1005,transportresistivity}. Subsequently, two transitions were detected by a recent electrical transport study in the single crystal sample, with transition temperatures of $\sim$ 110 K and 153 K, respectively\cite{Liu2022}. 
Most intriguingly, transport measurements show that $T_{\rm DW}$ decreases with increasing pressure and disappears around 6 GPa, where the superconductivity emerges\cite{polycrystalsol-gel,327CDWexperiments}. Such behavior suggests a competition between the possible DW and superconducting state, which is similar to that in the cuprate superconductors\cite{cuprateReview}. We also note that recent muon spin relaxation ($\mu$SR) study suggested that the transition temperature $T_{\rm DW}$ rises with the increase of pressure\cite{uSR2highpressure}. 
In any cases, an essential question arises regarding the existence of a DW order in La$_3$Ni$_2$O$_{7-\delta}$, and if it exists, whether it is a spin density wave (SDW) or a charge density wave (CDW). Or, could it be a stripe order, where the spin and charge orders are coupled, as observed in La$_{2-x}$Ba$_x$CuO$_4$\cite{LBCO}. 
Various experimental methods have been utilized to investigate the nature of the possible DW state, but the results are contradictory. The appearance of the internal magnetic field in La$_3$Ni$_2$O$_{7-\delta}$ below $T \sim$ 150 K observed by $\mu$SR\cite{uSR1shulei,uSR2highpressure} and nuclear magnetic resonance (NMR) \cite{327NMRwutao} point to a SDW order at low temperatures. This is also supported by the resonant inelastic X-ray scattering (RIXS), which reveals a quasi-static SDW order\cite{327magneticexcitations2}. In all of these studies, it was found that the magnetic moment of Ni was rather small, only $\sim$ 0.5 $\mu_B$/Ni. This implies that La$_3$Ni$_2$O$_{7-\delta}$ might differ from infinite-layer nickel oxide materials and does not present strong magnetic coupling\cite{CuiCPL}. The SDW could stem from the scattering between different sections of the Fermi surface. However, no long-range magnetic order was observed by neutron scattering (NS) measurements\cite{327magneticexcitations1}. Other NMR measurements conducted on the polycrystal sample argued that the broadening of the NMR spectra is more associated with the CDW order\cite{327NMR1,327NMR2}. The infrared spectroscopy investigation also supports the CDW order, but the CDW gap was found to be opened below  $T \sim$ 115 K\cite{opticalstudy}.  
The reason for these conflicting results can be attributed to the presence of multiple R-P phases in the La$_3$Ni$_2$O$_{7-\delta}$ samples, such as 4310 and 214 phases, as well as the intergrowth phases, as observed by the STEM and NQR measurements\cite{filamentarySC, 327LaPrNiOnature}. Especially, both SDW and CDW orders have been discovered by NS in La$_4$Ni$_3$O$_{10}$\cite{4310magneticorder}. Therefore, it is of urgency to find a way to evade the influence of other phases and merely acquire the microscopic information of the 327 phase. Meanwhile, the investigation of the properties of the DW order is also essential for understanding the role of electron correlations and the superconducting mechanism in La$_3$Ni$_2$O$_{7}$, as recent angle-resolved photoemission spectroscopy and optical studies have suggested that the electron correlation might be relatively strong\cite{ARPESElectronCorrelation,opticalstudy,opticalelectroniccorrelation1}.

In this study, we performed $^{139}$La-NQR to study the DW order in La$_3$Ni$_2$O$_{7-\delta}$. Taking advantage of the NQR measurement, we can get rid of the influence of other R-P phases in the sample and extract the microscopic information related to the La$_3$Ni$_2$O$_{7}$ phase\cite{327LaPrNiOnature}. Below the DW transition temperature $T_{\rm DW} \sim$ 153K, a distinct line splitting is observed in the $\pm$ 5/2 $\leftrightarrow$ $\pm$ 7/2 transition of the NQR resonance peak at the La(2) site, whereas only a line broadening is observed in the  $\pm$ 3/2 $\leftrightarrow$ $\pm$ 5/2 transition of the NQR resonance peak. Through further analysis of the spectra, we found that the line splitting is ascribed to the commensurate charge modulation, which is similar to the charge order in cuprate. A magnetic line broadening also emerges below $T_{\rm DW}$, indicating the formation of ordered moments. Meanwhile, a significant enhancement of the spin-lattice relaxation rate is further observed at $T_{\rm DW}$. All these findings suggest the simultaneous formation of spin and charge density wave order in La$_3$Ni$_2$O$_{7-\delta}$ at the same temperature.

The polycrystalline sample of La$_3$Ni$_2$O$_{7-\delta}$ was synthesized by sol-gel method\cite{polycrystalsol-gel}, and used in the earlier work\cite{327LaPrNiOnature}. The oxygen stoichiometry parameter $\delta \approx$  0.07 was determined by thermogravimetric analysis. $^{139}$La-NQR spectra were obtained by sweeping the frequency point by point, and integrating spin-echo intensity. The quantity of our sample used for NQR measurement is about 600 mg. The spin-lattice relaxation time $T_1$ was measured by saturation-recovery method. Because $\eta$ is close to zero, we ignore the influence of $\eta$ on the recovery curve. 

\begin{figure}[htbp]
\includegraphics[width=14cm]{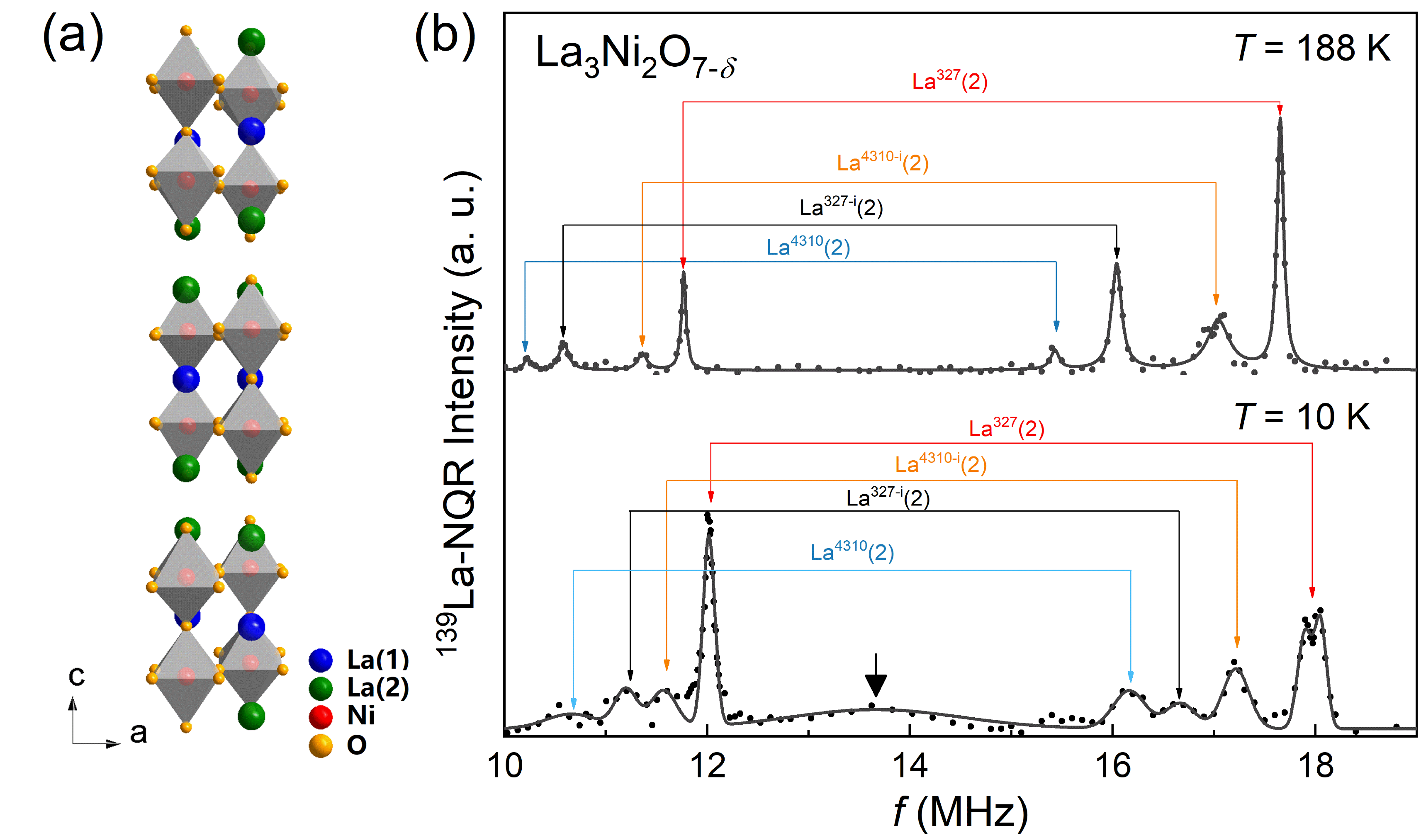}
\centering
\caption{(Color online) (a) The crystal structure of La$_3$Ni$_2$O$_{7}$ at ambient pressure. (b) The $^{139}$La-NQR spectra at 188 K and 10 K, respectively. There are two sets of resonance peaks, corresponding to  $\pm$ 5/2 $\leftrightarrow$ $\pm$ 7/2 and $\pm$ 3/2 $\leftrightarrow$ $\pm$ 5/2 transitions within the range of 10 MHz to 19 MHz. The two resonance peaks between the peaks of La$^{4310}$(2) and La$^{327}$(2) are from La$^{327-\rm i}$(2) and La$^{4310-\rm i}$(2) sites, which correspond to the La(2) sites from the intergrowth between 327 and 4310 phases (see Supplementary Fig. 1)\cite{Supplementary}. The solid arrow represents the broad peak corresponding to the La$_2$NiO$_4$ phase. At 188 K and 10 K, the spectra are respectively fitted by Lorentz and Gaussian functions (solid lines). 
}
\label{fullspectrum}
\end{figure}

In the crystal structure of La$_3$Ni$_2$O$_{7}$ shown in Fig.\ref{fullspectrum}(a), there are two types of La sites. Namely, La(1) sites are located inside the NiO$_2$ bilayer, and La(2) sites are outside the NiO$_2$ bilayer. The Hamiltonian caused by quadrupole interaction at zero field is $\mathcal{H_Q} = \frac{e^2qQ}{4I(2I-1)}[(3I_z^2-I^2)+\frac{1}{2} \eta\left(I_{+}^{2}+I_{-}^{2}\right)]$ \cite{Abragam}. $eq$ = $V_{ZZ}$ = $\frac{ \partial^2{V}}{\partial {Z^2}}$ is the Electric Field Gradient (EFG) along the principle axis and $V$ is the electric potential, $Q$ is the nuclear quadrupole moment, and $\eta$ = $\lvert$$(V_{XX}$ - $V_{YY})$/$V_{ZZ}$$\lvert$ is the asymmetry parameter of EFG. The nuclear quadrupole resonance frequency $\nu_{\rm Q}$ is defined as $\frac{3e^2qQ}{2I(2I-1)h}$. For the $^{139}$La nucleus with spin 7/2, the NQR spectrum should have three resonance peaks corresponding to  $\pm$ 1/2 $\leftrightarrow$ $\pm$ 3/2,  $\pm$ 3/2 $\leftrightarrow$ $\pm$ 5/2 and  $\pm$ 5/2 $\leftrightarrow$ $\pm$ 7/2 transitions. Thus, a total of six sets of peaks should be observed in the $^{139}$La-NQR spectrum for La$_3$Ni$_2$O$_{7}$. However, the $\nu_{\rm Q}$ of La(1) is only 2.2 MHz\cite{327LaPrNiOnature,327NMRwutao,327NMR1,327NMR2}. So we only observe two sets of resonance peaks in the NQR spectra corresponding to the La(2) site between 10 to 19 MHz as shown in Fig. \ref{fullspectrum}(b). At $T$ = 188 K, each set exhibits four peaks in the La(2) NQR spectrum (Fig. \ref{fullspectrum}(b), top), indicating the presence of four distinct La(2) sites. Based on the previous STEM and NQR study\cite{327LaPrNiOnature}, we can ascribe the lowest-frequency peak to the 4310 phase. The two intermediate peaks originate from two sites within the intergrowth between the 327 and 4310 phases (see Supplementary Fig. 1\cite{Supplementary}). Nevertheless, in the previous NQR study\cite{327LaPrNiOnature}, 1/$T_1$$T$ measurements were not carried out, and the lower-frequency peak was assumed to be La$^{4310-i}$(2). From the temperature dependence of the 1/$T_1$$T$ results at different peaks (see Supplementary Fig. 5)\cite{Supplementary}, we are now able to assign the resonance peaks, from low to high frequency, to the  La$^{4310}$(2), La$^{\rm 327-i}$(2), La$^{\rm 4310-i}$(2), and La$^{327}$(2) sites.

%as well as the temperature dependence of the 1/$T_1$$T$ results at different peaks (see supplementary Fig. 5 ), we attribute the resonance peaks from low to high frequency to the La$^{4310}$(2), La$^{\rm 327-i}$(2), La$^{\rm 4310-i}$(2), and La$^{327}$(2) sites. Here, La$^{\rm 327-i}$(2) and La$^{\rm 4310-i}$(2) are two sites from the intergrowth between 327 and 4310 phases(see Supplementary Fig. 1)\cite{Supplementary}.
%From the previous STEM and NQR study\cite{327LaPrNiOnature}, the resonance peaks from the lowest and highest frequency are from La$^{4310}$(2) and La$^{327}$(2) sites, which respectively correspond to the La(2) site in the La$_4$Ni$_3$O$_{10}$ phase and La$_3$Ni$_2$O$_{7}$ phase. 

\begin{figure}[htbp]
\includegraphics[width=11 cm]{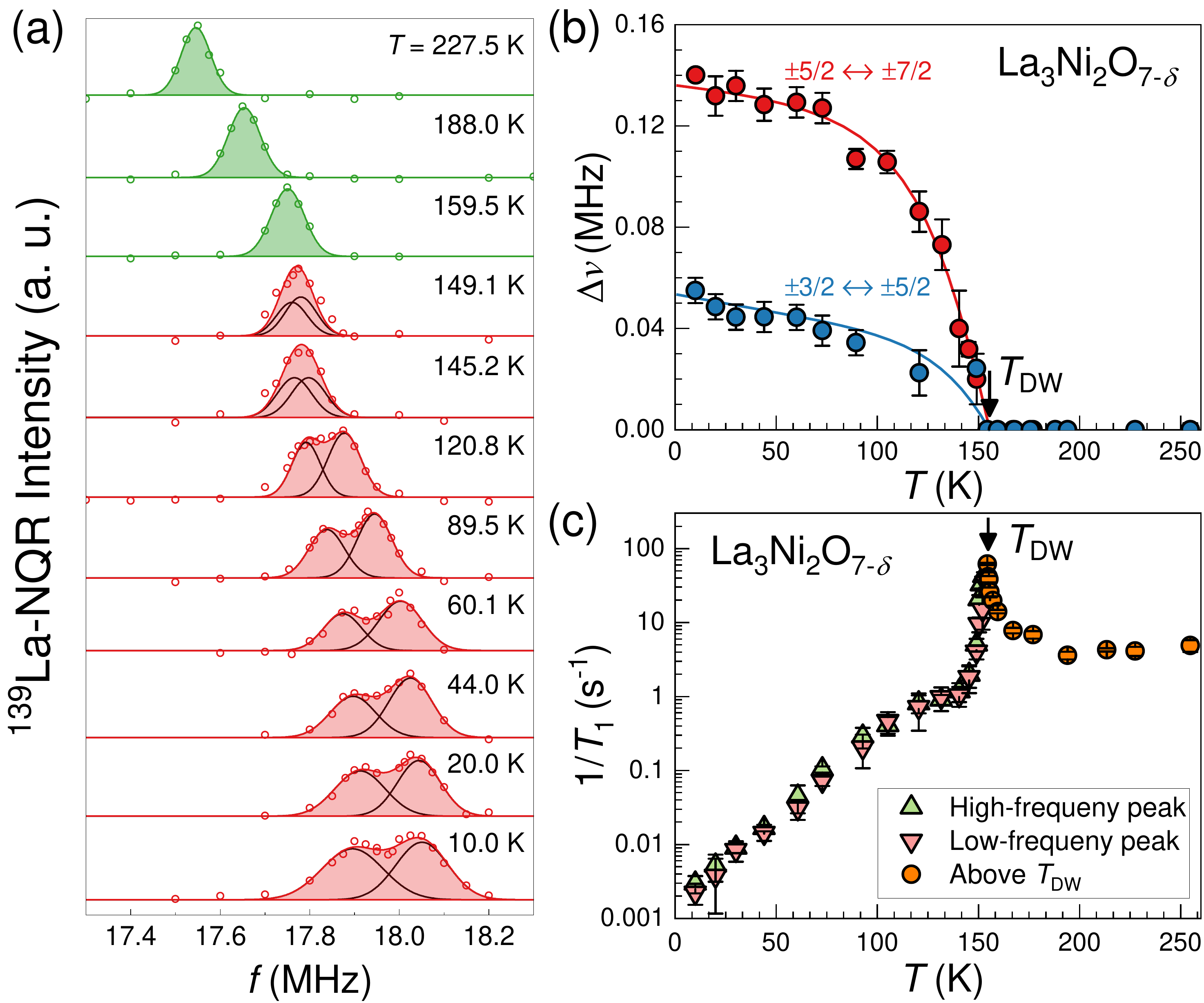}

\caption{(Color online) (a) The temperature dependence of the  La$^{327}$(2) NQR spectra corresponding to the $\pm 5/2 \leftrightarrow \pm 7/2$ transition. %As temperature decreases, the shift to high frequency. 
Above $T_\textrm{DW}$, the NQR spectra are fitting by one Gaussian function. Below $T_\textrm{DW}$, the NQR spectra are fitting by two Gaussian  functions.  (b) Temperature dependence of the NQR line splitting of both $\pm$ 3/2 $\leftrightarrow$ $\pm$ 5/2 and  $\pm$ 5/2 $\leftrightarrow$ $\pm$ 7/2 transitions obtained from the fitting of the NQR spectra\cite{Supplementary}. Below the DW phase transition, the NQR line splitting $\Delta \nu$ increases rapidly with decreasing temperature. %The error bars are the s. d. in fitting the spectra. 
(c) Temperature dependence of 1/$T_1$ measured at the $\pm$ 5/2 $\leftrightarrow$ $\pm$ 7/2 transition. The black arrows represent $T_{\rm DW}$. %The error bar of $1/T_1$ is the s. d. in fitting the recovery curve.
}
\label{Tdependentspectrum}
\end{figure}

With decreasing temperature below $T_{\rm DW}$, a clear splitting appears at La$^{327}$(2) site of the $\pm$ 5/2 $\leftrightarrow$ $\pm$ 7/2 transition, while line broadening can be seen at other sites(see Fig. \ref{fullspectrum}(b), bottom). A broad peak at $\sim$ 13 MHz is observed below $T$ = 30 K\cite{Supplementary}. Considering that the 214 phase was detected by STEM in the polycrystal sample\cite{327LaPrNiOnature}, and a similar broad peak was observed in the previous NQR study\cite{La2NiO4AFM}, we ascribe this broad peak to the 214 phase.
In order to obtain the intrinsic information from the 327 phase, we thus focus on the La$^{327}$(2) site. Figure \ref{Tdependentspectrum}(a) presents the temperature dependence of the resonance peak of the La$^{327}$(2) site corresponding to the ${\pm 5/2 \leftrightarrow \pm 7/2}$ transition. Below $T_{\rm DW} \sim$ 153 K, the peak begins to broaden and further splits at low temperatures. For the La$^{327}$(2) NQR spectra corresponding to the ${\pm 3/2 \leftrightarrow \pm 5/2}$ transition, only the line broadening can be observed down to 10 K\cite{Supplementary}. Then we fit the La$^{327}$(2) spectra corresponding to ${\pm 3/2 \leftrightarrow \pm 5/2}$ and ${\pm 5/2 \leftrightarrow \pm 7/2}$ transitions with two Gaussian functions below $T_{\rm DW}$\cite{Supplementary}, and obtain the interval of the two splitting lines $\Delta\nu$ as shown in Fig. \ref{Tdependentspectrum}(b). A rapid increase of $\Delta\nu$ just below $T_{\rm DW}$ is seen for both ${\pm 3/2 \leftrightarrow \pm 5/2}$ and ${\pm 5/2 \leftrightarrow \pm 7/2}$ transitions. Meanwhile, a significant enhancement of 1/$T_1$ with the decrease of temperature down to $T_{\rm DW}$ is also observed as shown in Fig. \ref{Tdependentspectrum}(c). The increase of 1/$T_1$ from $T \sim$ 180 K down to $T_{\rm DW}$ is due to the fluctuations of the possible DW order. Below $T_{\rm DW}$, 1/$T_1$ is measured at both splitting lines, and both are found to decrease with the decrease of temperature due to the gap opening. All these indicate the appearance of a DW phase transition at $T_{\rm DW}$. The $1/T_1T$ below 50 K decreases to $1\%$ of that at high temperatures above $T_{\rm DW}$, suggesting that most part of the Fermi surface is gapped. Furthermore, we also note that the temperature dependence of 1/$T_1$$T$ results at the La$^{\rm 4310-i}$(2) site suggests that the $T_{\rm DW} \sim$ 133 K of the intergrowth 4310 phase is still close to the transition temperature of the pure 4310 phase identified in previous NS studies\cite{4310magneticorder,Supplementary}. This explains why two DW transitions can occasionally be detected in the La$_3$Ni$_2$O$_7$ sample through electrical resistivity measurements\cite{Liu2022}. 

\begin{figure}[htbp]
\includegraphics[width=15 cm]{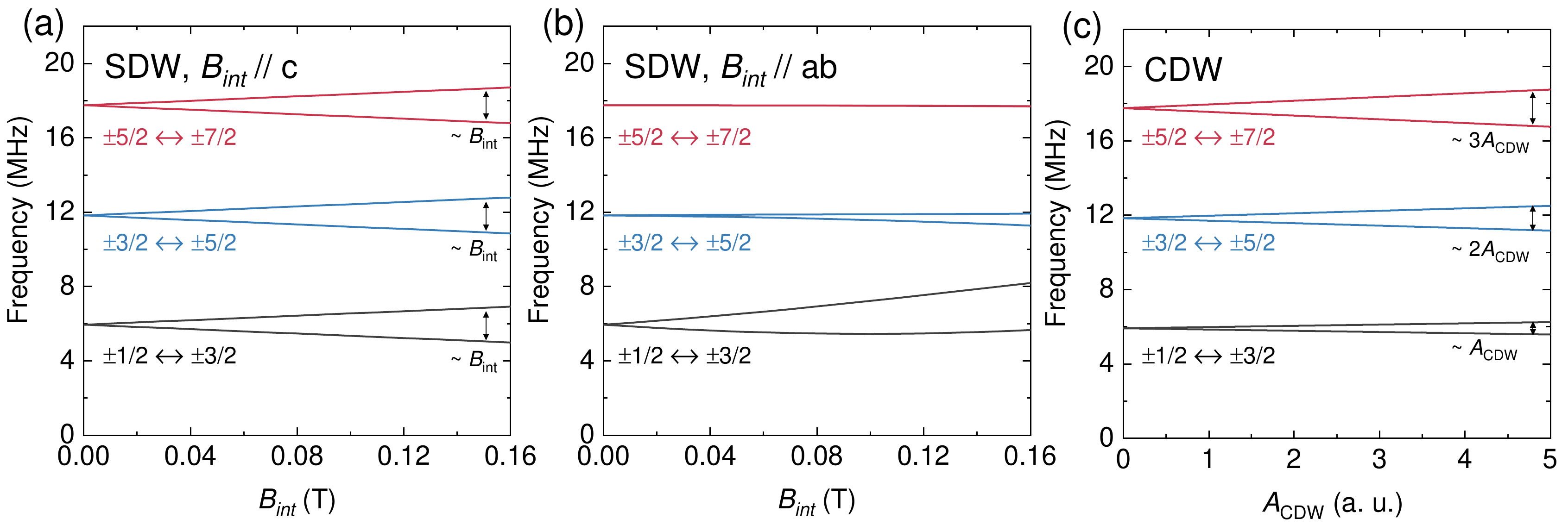}

\caption{(Color online) The calculated evolution of the NQR lines in the presence of the internal field along the $c$ axis (a) and $ab$ plane (b) for the La(2) site. (c) The calculated evolution of the NQR lines in the presence of the CDW order as functions of the amplitude of the charge modulation. For all calculations, $\nu_Q$ and $\eta$ are taken as 5.92 MHz and 0.04, respectively.
}
\label{cal}
\end{figure}

The observed line splitting can be due to either an emergence of an internal magnetic field or a modulation of the charge density. As previously indicated by NMR measurements, the principal axis of the EFG for the La site is along the $c$-axis\cite{327NMRwutao}. Then, in the case where the internal magnetic field $B_{int}$ is along the $c$ axis due to the SDW order, the NQR line splitting should be $2\gamma B_{int}$ for all transitions (see Fig. \ref{cal}(a)), where $\gamma$ = 6.0146 MHz/T is the nuclear gyromagnetic ratio of $^{139}$La.  We would expect to observe that $\Delta\nu$ at both ${\pm 3/2 \leftrightarrow \pm 5/2}$ and ${\pm 5/2 \leftrightarrow \pm 7/2}$ transitions is identical. 
For the case of the internal magnetic field $B_{int}$ along the $ab$ plane from the SDW order as suggested by previous NMR and RIXS studies\cite{327NMRwutao,327magneticexcitations2}, the NQR line splitting at the ${\pm 5/2 \leftrightarrow \pm 7/2}$ transition should be barely seen and much smaller than the line splitting at the ${\pm 3/2 \leftrightarrow \pm 5/2}$ transition (see Fig. \ref{cal}(b)). Therefore, our results imply that the NQR line splitting is unlikely originate from the SDW order. 

For the case of the charge modulation, the NQR line splitting should be proportional to the NQR frequency. We would expect to observe that $\Delta\nu$ at ${\pm 5/2 \leftrightarrow \pm 7/2}$ transition should be 1.5 times larger than that of ${\pm 3/2 \leftrightarrow \pm 5/2}$ (see Fig. \ref{cal}(c)), when the $\eta$ of two La(2) sites are the same. As shown in Fig. \ref{Tdependentspectrum}(b), $\Delta\nu$ of the ${\pm 5/2 \leftrightarrow \pm 7/2}$ transition is about 2 times larger than that of the ${\pm 3/2 \leftrightarrow \pm 5/2}$, which is slightly larger than expected value 1.5.
It is possible that the EFG asymmetry parameter $\eta$ of the two splitting lines is not identical, or that there exists a very small internal magnetic field along the $ab$ plane at La(2) sites. All these could lead to the observed $\Delta\nu$ ratio between ${\pm 5/2 \leftrightarrow \pm 7/2}$ and ${\pm 3/2 \leftrightarrow \pm 5/2}$ being slightly larger than the expected value 1.5. In any case, our observation still indicates that the observed splitting is mainly attributed to the charge modulation resulting from a CDW order below $T_{\rm DW}$. Such a behavior is similar to the previously observation in cuprates\cite{CDWinYBCO,Kawasaki2017} and kagome metals\cite{npjluo}. To obtain the precise parameters from the NQR measurements, narrower lines from high-quality single crystals and the measurements on the ${\pm 1/2 \leftrightarrow \pm 3/2}$ transition are required in the future.  
%There are certain factors that can slightly influence the $\Delta\nu$ ratio between ${\pm 5/2 \leftrightarrow \pm 7/2}$ and ${\pm 3/2 \leftrightarrow \pm 5/2}$ transitions, such as the variation of $\eta$ in the CDW state, 
%Meanwhile, our results also implies 

\begin{figure}[htbp]
\includegraphics[width=13cm]{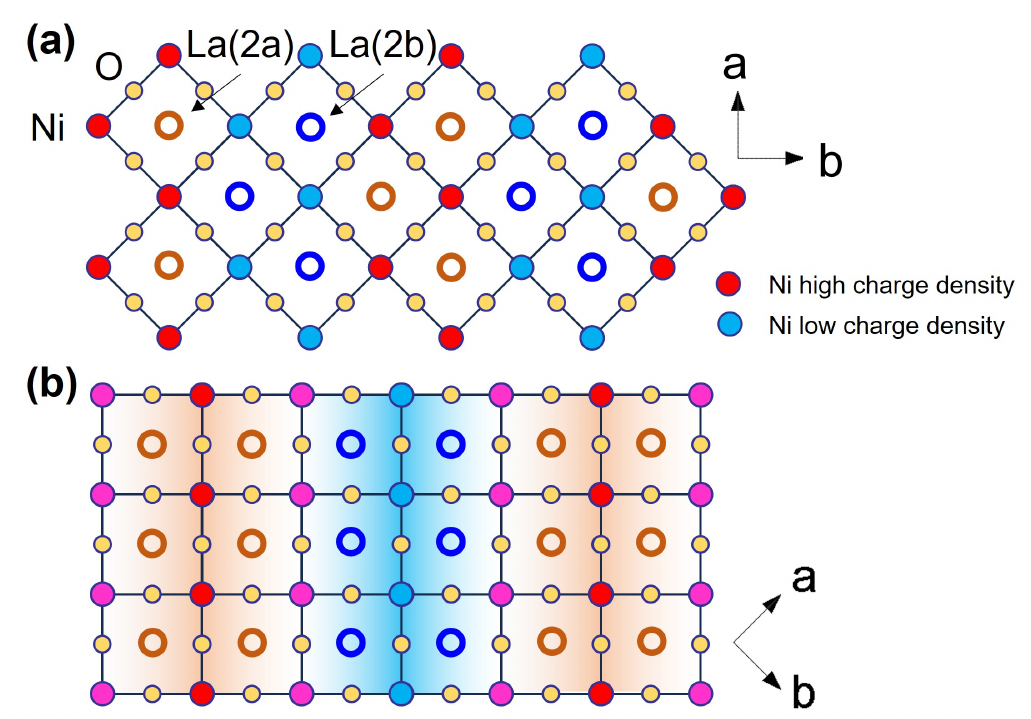}
\centering
\caption{(Color online) Illustration of possible CDW patterns below $T_{\rm DW}$. The double charge stripe order and the unidirectional CDW along the Ni-O-Ni bond direction compatible with the NQR spectra are shown in (a) and (b). The red (blue) solid circle represents Ni with high (low) charge density. The open circles represent two La(2) sites, marked as La(2a) and La(2b). The light orange and light blue shades indicate high charge density and low charge density area in (b).}
\label{CDWstructure}
\end{figure}

For the one-dimensional (1D) incommensurate CDW order\cite{Blinc2002,npjfeng}, the NMR spectrum is expected to have two peaks of equal intensities, and most importantly, a continuum between the two peaks. In the cases of 2D or 3D, if the amplitude of the charge modulation along one direction is much stronger than the others, the NMR spectrum should be similar to that of the 1D incommensurate CDW order. Otherwise, the spectrum should only have one symmetric peak, which is not consistent with our observation. Considering a line splitting in our spectra within the CDW state, we then simulate the NQR spectra at 10 K (see Supplementary Fig. 8)\cite{Supplementary}. We find that the fitting of two Gaussian functions is more consistent with our experimental results than the simulation of 1D-incommensurate CDW order, suggesting that the charge modulation is commensurate.
Then the two peaks of $^{139}$La-NQR spectra indicate that there are two types of La(2) sites in the CDW state. Considering that La(2) is at the center of four nearby Ni atoms, we propose two possible charge modulation patterns as shown in Fig. \ref{CDWstructure}. The first possible CDW pattern is the double charge stripe order (see Fig. \ref{CDWstructure}(a)). In this CDW state, there are two types of La(2) sites, arranged in a Zig-Zag chain. This CDW pattern is consistent with the CDW-Z1 mode proposed in a previous theoretical work\cite{327CDWtheory1}, where the CDW was proposed to be induced by the tendency of structural distortion due to the four imaginary modes existing in the phonon spectrum at ambient pressure. Another possible CDW pattern is shown in Fig. \ref{CDWstructure}(b), in which the stripe type charge modulation is along the Ni-O-Ni bond direction. In this case, there are also two types of La(2) sites. This type of CDW is similar to long-range commensurate CDW order in Bi$_2$Sr$_2$CaCu$_2$O$_{8+x}$\cite{Mesaros2016} and YBa$_2$Cu$_3$O$_y$\cite{Igor2021}.  In any case, our results suggest that the CDW order is commensurate and unidirectional. In principle, additional diffraction peaks should be observable by X-ray scattering due to charge modulation, however, this was not observed so far\cite{327magneticexcitations2}. This might be attributed to the fact that X-ray scattering, as a bulk measurement, is influenced by the presence of other structural phases, such as 4310, 214 and other intergrowth phases, which would blur the signature of the charge modulation.

\begin{figure}[htbp]
\includegraphics[width=10cm]{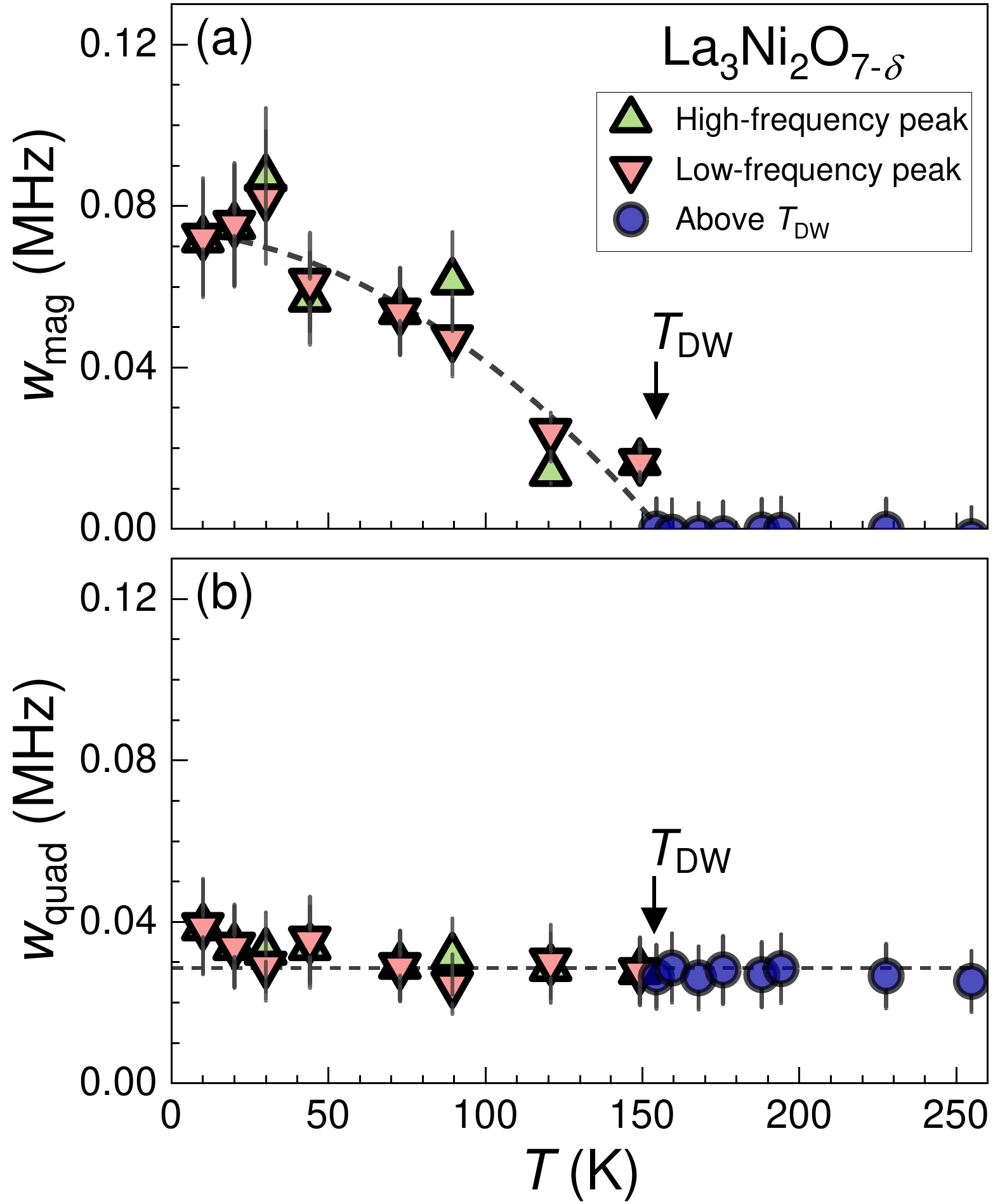}
\centering
\caption{(Color online) The magnetic (a) and quadrupole (b) contribution to the FWHM extracted from the total FWHM\cite{Supplementary},  including both high- and low-frequency peaks of two split peaks below $T_{\rm DW}$ and a single peak above $T_{\rm DW}$. The black arrows represent $T_{\rm DW}$. The dashed lines are the guides to the eye. 
%Above $T_{\rm DW}$, the magnetic contribution maybe imaginary number owing to poor signal to noise, we set the error bar as absolute value of the imaginary number. %The error bar below $T_{\rm DW}$  is transmitted from fitting error of $\pm$ 3/2 $\leftrightarrow$ $\pm$ 5/2 and $\pm$ 5/2 $\leftrightarrow$ $\pm$ 7/2. The error bar is the s. d. in fitting the spectra.
}
\label{FWHM}
\end{figure}

We can further obtain information in the DW state from the temperature dependence of the full width at half maximum (FWHM),   including both high- and low-frequency peaks of two split peaks below $T_{\rm DW}$ and a single peak above $T_{\rm DW}$\cite{Supplementary}. Above $T_{\rm DW}$, the FWHM is independent of temperature, but it starts to increase below $T_{\rm DW}$ for both ${\pm 5/2 \leftrightarrow \pm 7/2}$ and ${\pm 3/2 \leftrightarrow \pm 5/2}$ transitions. In principle, the broadening of the NQR spectrum can be attributed to both quadrupole and magnetic contributions. The FWHM can be expressed by an ad hoc formula, $w_{\rm tot}$ = $\sqrt{(m \cdot w_{\rm quad})^2 + (w_{\rm mag})^2}$. The $w_{\rm quad}$ and $w_{\rm mag}$ are the contributions of the quadrupole and magnetic contributions to the line broadening. The value of $m$ is 2 for the $\pm 3/2 \leftrightarrow \pm 5/2$ transition and 3 for the $\pm 5/2 \leftrightarrow \pm 7/2$ transition, respectively. Therefore, we can deduce the magnetic contribution $w_{\rm mag}$ and the quadrupole contribution $w_{\rm quad}$ from the total FWHM $w_{\rm tot}$, as shown in Fig.~\ref{FWHM}(a) and (b), respectively. Above $T_{\rm DW}$, the magnetic broadening $w_{\rm mag}$ is nearly zero. However, $w_{\rm mag}$ of both high- and low-frequency peaks of two split peaks starts to increase below $T_{\rm DW}$, becoming the dominant contribution to the total FWHM at low temperatures. In contrast, the quadrupole contribution $w_{\rm quad}$ remains almost temperature-independent within the DW state. The emergence of magnetic broadening at zero field indicates the formation of a magnetic ordered moment associated with the spin density wave in La$_3$Ni$_2$O$_7$. 
Meanwhile, we observe that $1/T_1$ is $T$-independent at high temperatures above $T \sim$ 180 K but increases by nearly two orders of magnitude at $T_{\rm DW}$ in La$_3$Ni$_2$O$_{7}$ (see Fig.~\ref{Tdependentspectrum}(c)).  Such behavior contrasts with the situation in kagome metals where $1/T_1$ decreases as the temperature drops down to the CDW transition temperature \cite{npjluo}. 
The $T$-independent 1/$T_1$ at high temperatures is consistent with the existence of local magnetic moments\cite{Moriya1956}, and resembles previous observations in La$_{2-x}$Sr$_x$CuO$_4$ near the spin ordering transition\cite{LaSrCuO4NMR,HiddenmagnetisminLaSrCuO4}.
All these indicate that the enhancement of $1/T_1$ originates from spin fluctuations, providing further evidence for the presence of SDW order below $T_{\rm DW}$. 
%All these indicate that both the CDW and SDW coexist below $T_{\rm DW}$ in La$_3$Ni$_2$O$_{7}$.
%Therefore, the substantial increase of 1/$T_1$ is most likely attributed to the strong spin fluctuations from the SDW order. All these indicate that both the CDW and SDW coexist below $T_{\rm DW}$ in La$_3$Ni$_2$O$_{7}$. 

In NQR measurements, since no external magnetic field is applied and polycrystalline samples are used, the line broadening might arise from the distribution of magnetic moments in any direction due to the oxygen content distribution within the sample. Consequently,  we can not provide more information about the structure of the SDW order. However, we note that the observed internal field $B_{int}$ at the La(2) site can be estimated from the magnetic line broadening as $B_{int} \sim$ 0.005 T, which is close to the half of the $B_{int}$ at the La(1) site\cite{327NMRwutao}. Considering that La(1) site is between two NiO$_2$ layers and La(2) site is only close to one NiO$_2$ layer, the observed $B_{int}$ value suggests that our observation is consistent with the NMR measurements\cite{327NMRwutao}.Moreover, the small $B_{int}$ led to only broadening being observed on the NQR spectra, rather than splitting. The double spin stripe is proposed for the structure of the SDW order by previous NMR and RIXS study\cite{327NMRwutao,327magneticexcitations2}. Our results further imply that the double charge stripe could be a possible structure of the CDW order. Therefore, in combination with the previous study on the SDW order, all these suggest that the CDW and SDW might be coupled in La$_3$Ni$_2$O$_7$, similar to the stripe order in La$_{2-x}$Ba$_x$CuO$_4$\cite{LBCO}. Meanwhile, the commensurate nature of the CDW order is also similar to that of the cuprates\cite{Mesaros2016,Igor2021}. 
However, there is a striking difference between Ni-based compounds and the cuprates. In the latter, the undoped parent compound is a Mott insulator, but the former is not, with an electrical conductivity worse than the kagome metal though\cite{Wilson2024}. In the parent cuprates such as La$_2$CuO$_4$, the internal magnetic field due to the antiferromagnetic interaction is huge, with the hyperfine field about 0.1 T at La-site\cite{Kitaoka1988}, which is much larger than we found here. Although CDW is also present in the cuprates,  except a particular doping rate 1/8, the order is only short range which can be stabilized to become long range only after applying strong magnetic field\cite{CDWinYBCO,Kawasaki2017} or uniaxial strain\cite{Kim2018,kawasaki2024}. By contrast, the CDW long-range order is robust already in the undoped nickelates. Furthermore, we notice that La$_3$Ni$_2$O$_7$ bears similarities to the trilayer compound La$_4$Ni$_3$O$_{10}$, in which both CDW and SDW occur simultaneously. One possibility is that the SDW emerges from Fermi surface nesting\cite{La3Ni2O7Electronicandmagneticstructure,La3Ni2O7Electronicinstability}. Subsequently, spin-lattice coupling might concurrently induce charge ordering, offering a potential mechanism for the coexistence of charge and magnetic order\cite{La3Ni2O7SDWambient}. Additionally, it is also possible that La$_3$Ni$_2$O$_7$ is similar to La$_4$Ni$_3$O$_8$\cite{La4Ni3O8CDWandSDW}, where spin and charge stripes are strongly coupled\cite{La3Ni2O7Peierlsinstability}, and neither order parameter is secondary. In any case, our results thus provide new perspectives for the study of high-temperature superconductors and can stimulate more theoretical and experimental investigations.

%\textcolor[rgb]{1.00,0.00,0.00}{suggesting that its origin could be related to its strong electronic correlations. }

%which was also found by $^{139}$La-NMR, $\mu$SR and RIXS measurements\cite{327NMRwutao,uSR1shulei,uSR2highpressure,327magneticexcitations2}.

%The linewidth caused by quadrupole interaction for $f_{\pm 5/2 \leftrightarrow \pm 7/2}$ and $f_{\pm 3/2 \leftrightarrow \pm 5/2}$ should be proportional to 3:2. However, the linewidth caused by magnetic interaction for the two transition is same, namely, 1:1.

%Below $T_{\rm DW}$, $1/T_1$ decreases rapidly, implicating that a energy gap opens. Through a exponential function fitting of $1/T_1$ data, we get the gap size is 32.7 meV, which is close to 70 meV obtained by ultrafast optical spectroscopy \cite{f327gapsize}. The $1/T_1T$ below 50K decrease to $1\%$ than that at high temperature, meaning that most of the fermi surfaces are gapped.

%\subsection{The structure of CDW order}

%\subsection{The accompanying spin density wave order}

%\section{Conclusion}
In summary, we have performed $^{139}$La-NQR to eliminate the influence of other R-P phases and obtain the microscopic information related to the DW transition in La$_3$Ni$_2$O$_{7-\delta}$. Below the DW transition temperature $T_{\rm DW} \sim$ 153 K, a distinct line splitting is observed in the $\pm$ 5/2 $\leftrightarrow$ $\pm$ 7/2 transition of the NQR resonance peak at the La(2) site, indicating the unidirectional CDW order. Meanwhile, 1/$T_1$ is $T$-independent at high temperatures, but is largely enhanced towards $T_{\rm DW}$ upon cooling, with an emergence of magnetic broadening below $T_{\rm DW}$, pointing to the formation of magnetic ordered moments in the DW state. All these findings indicate the formation of CDW and SDW orders in La$_3$Ni$_2$O$_{7-\delta}$ simultaneously. Our results elucidate the nature of the DW order, thereby furnishing crucial information for understanding the electronic correlations in Ni-based high-temperature superconductors.

\vspace{0.5cm}

\begin{acknowledgments}
This work was supported by the National Key Research and Development Projects of China (Grants No. 2023YFA1406103, No. 2024YFA1611302, No. 2024YFA1409200 and No. 2022YFA1403402), the National Natural Science Foundation of China (Grants No. 12374142, No. 12304170, No. 12025408, No. 12404179 and No. U23A6003), Beijing National Laboratory for Condensed Matter Physics (Grant No. 2024BNLCMPKF005) and the Chinese Academy of Sciences President’s International Fellowship Initiative (Grant No. 2024PG0003). This work was supported by the Synergetic Extreme Condition User Facility (SECUF, https://cstr.cn/31123.02.SECUF).
\end{acknowledgments}


%\vspace{0.5cm}
%\textbf{Author contributions}

%The crystal sample was grown by W. Gang and J.G.Cheng. The NQR measurements were performed by J. Luo, J. Feng, J. Dou, J. Yang, A. F. Fang and R. Zhou. R. Zhou and G.-q. Zheng wrote the manuscript with inputs from J. Luo. All authors have discussed the results and the interpretation.

\end{document}